\documentclass[manuscript,usenatbyb]{aastex}

\slugcomment{Not to appear in Nonlearned J., 45.}

\shorttitle{Initial spin periods of neutron stas}
\shortauthors{Popov, Turolla}

\begin{document}


\title{Initial spin periods of neutron stars in supernova remnants}


\author{S.B. Popov\altaffilmark{1}}
\affil{Sternberg Astronomical Institute,
Lomonosov Moscow State University, Russia}
\email{sergepolar@gmail.com}

\and

\author{R. Turolla}
\affil{Department of Physics and Astronomy, University of Padova,
Italy}
\affil{Mullard Space Science Laboratory, University College London, UK}


\altaffiltext{1}{Visiting Scientist, University of Padova}


\begin{abstract}
We present estimates of initial spin periods, $P_0$, for radio pulsars
associated with supernova remnants. By using the published data on 30
objects, we were able to derive a reliable estimate for the initial spin
period, assuming standard magneto-dipole spin-down (braking index $n=3$),
in many cases.
Our set of estimates is still not sufficient to infer the exact shape of
the initial period distribution. However, we show that a gaussian
distribution with mean and deviation $\sim 0.1$~s is consistent
with our results, while flat, wide distributions and very narrow ones are
disfavored.
\end{abstract}






\section{Introduction}

Studies of neutron star (NS) properties and evolution is a
fruitful field of research uniting astrophysics and fundamental
physics (see a general review and references, for example, in the
introductory part of the book \citealt{hyp2007} and a shorter review
in \citealt{lp2004}). In the map of this research area quantum
chromodynamics is linked to observations of cooling NSs; general
relativity with pulse profiles of X-ray sources; strong field
electrodynamics with astrophysics of radiopulsars and magnetars,
and so on. Despite significant progress, some parts of this
picture are still obscure.  One of them is related to NSs initial
parameters. The knowledge of initial properties of compact objects
is of crucial importance not only for understanding their
evolution and modelling their observational appearance, but also
because the parameters of newborn NSs carry the imprint of the still
poorly understood supernova (SN) explosion mechanism.

For a NS the main initial parameters are the mass ($M$), dipolar
external magnetic field ($B$), spin period ($P$) and spatial
velocity ($v$). In addition, the angle between the spin and
magnetic axis, the initial configuration of the crustal magnetic field
(including a toroidal component), and other parameters are
important for the thermal and/or  magneto-rotational evolution and
for fixing the different observational manifestations of NSs.

There is strong, ongoing effort by many research groups in the
attempt to advance our knowledge of NSs initial parameter
distributions. This can be done by different techniques. Direct
observations, unfortunately, are impossible, but observational
data can be used together with assumptions about the NS evolution to
estimate a posteriori the initial parameters for individual
objects.
Another approach is related to population synthesis
\citep{pp2007}. In this case an evolutionary scenario is applied to
follow numerically a large sample of sources, selection effects are
taken into account, and results are confronted with
observational distributions. Finally, initial parameters can be
calculated in the framework of some theoretical model (for
example, models of SN explosion).

Population synthesis modelling was used to study the
initial properties of radiopulsars (PSRs) by \cite{fgkaspi}, who
considered several possible initial distributions of NS parameters in
the case of standard PSRs. Another approach was used in
\cite{popov2010}, where in the framework of a decaying magnetic
field, several types of isolated NSs were studied, with a focus on
the properties of the initial magnetic field.
A direct numerical model is used, for example, to derive
initial mass distribution of compact objects \cite[e.g.][]{fryer2011}.
Another example is related to the calculation of the kick (recoil)
velocity of newborn compact objects \citep{scheck2006}.

In this note we address the problem of determining the initial
spin period  $P_0$ of NSs. This is done by using data on the ``true
age'' of NSs, as obtained by observations of associated SN remnants
(SNRs) and applying a standard model of spin evolution. In the
next section we present our sample and results. Then we provide a
discussion of several issues related to our study, and, finally,
conclude.

\section{Sample and initial periods}

We collected a sample of 30 NSs with well-measured $P$ and $\dot P$,
associated with SNRs or/and pulsar wind nebulae (see Table 1 for a
complete list). For all of them there are estimates of
SNR ages based on different assumptions
(in the case of the Crab nebula and, probably, G11.2-0.3
historical ages can be used, see \citealt{green2003}).
Period and period derivatives are taken from the ATNF catalogue
\citep{atnf}\footnote{The on-line catalogue is available at
http://www.atnf.csiro.au/people/pulsar/psrcat/}.

\begin{figure}
\epsscale{.99}
\plotone{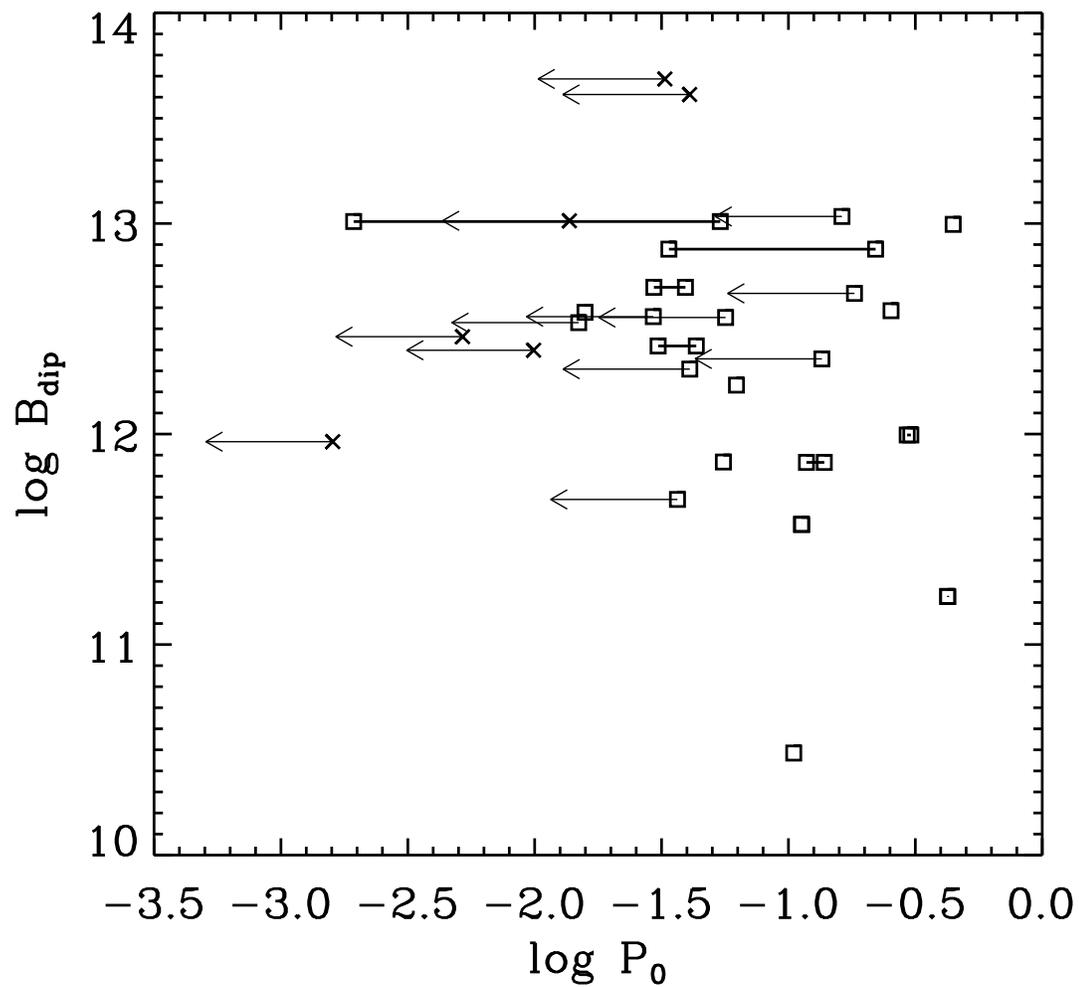}
\caption{Magnetic fields and initial spin periods for 22 PSRs (squares).
In several cases
only upper limits are available (shown with arrows).
For the group of six pulsars with SNR
ages in correspondence with spin-down ages we use different symbols
(crosses) and assume that $P_0<0.1P$, where $P$ is the present period.
All values are estimated for standard
magneto-dipole losses in the case of constant field and angle between
spin and magnetic axis.
\label{fig1}}
\end{figure}

In 6 cases the independently estimated
SNR age, $\tau_{\mathrm{SNR}}$, is consistent with the
spin-down age, $\tau_{\mathrm{sd}}=P/2\dot P$, calculated for
zero initial spin periods and braking index $n=3$
($n\equiv \nu \ddot \nu /\dot \nu ^2,$ where $\nu=1/P$ is the
spin frequency), and so no
estimate of $P_0$ is  possible in the standard picture,
except that it has to be $P_0\ll P$. Note, that this does not formally
mean
that $P_0\simeq 0$. The initial period for these sources can range from
a few
ms up to tens of ms (as in, for example, PSR J1119-6127 and PSR
J1846-0258). In plotting the estimated  initial spin period for
these sources (Fig. \ref{fig1}) we assume $P_0<0.1P$.

In three cases only upper or lower limits on a SNR age are available.
For 15 SNRs age ranges are reported in the literature. For most of them it
is possible to obtain a range of
initial spin periods and for some the resulting ranges for
$P_0$ are narrow. In several cases
$\tau_\mathrm{sd}$ is in between the upper and lower limit of
$\tau_\mathrm{SNR}$. Correspondingly, only an upper limit on $P_0$
can be derived.
Finally, in 6 cases the SNR ages are relatively well
determined, and in one case the uncertainty is tiny in
comparison with the spin-down age of the pulsar.
This allows us to obtain firm estimates of $P_0$ for a
given braking index.

Initial spin periods are estimated from the usual expression

\begin{equation}
P_0=\left[P^{n-1}-(n-1)\dot P P^{n-2} \tau_\mathrm{SNR} \right]^{1/(n-1)},
\end{equation}
where $P$ and $\dot P$ are the present values of the spin period and
period derivative, and $n$ is the braking index. For $n=3$ the
equation reduces to the usual magneto-dipole formula with
$B\propto \sqrt{P \dot P}$. These are the values presented in
Table 2 and Fig. 1. The field is assumed to be constant during the
lifetime of a NS, as well as the angle between spin and magnetic
axes. This assumption is well justified, as all sources are young
and do not possess strong magnetic fields. So, it is very unlikely that
field decay could  play
a major role in their evolution. Our estimates are in
good correspondence with those provided by \cite{migl2002}.
However, our list is significantly broader, although it is based
on less detailed observational data.

\subsection{Notes on individual objects}

{\bf N157B.} The source is situated in the Large Magellanic Cloud.
The pulsar was discovered by \cite{n157b} and the age of the
associated Crab-like
SNR is estimated in \cite{wg1998} as $\sim$5000 yrs basing on X-ray data
(ROSAT and
ASCA). Uncertainties in the age are not reported. The SNR age is in good
correspondence with the characteristic age of the pulsar.

{\bf G292.2-0.5.}  Both for free expansion and Sedov stage
\cite{g292.2-0.5} conclude that the age of the nebula can be  in
correspondence with the radio pulsar characteristic age.

{\bf G0.9+0.1.} This is a composite SNR. Age estimates reported by
\cite{g0.9+0.1} and \cite{porquet2003} are not precise, a few thousand
years. This is compatible with the characteristic age of the pulsar.

{\bf G359.23-0.82.} This is a pulsar wind nebula. There are no reliable
age estimates for this object which can be used for deriving the radio
pulsar initial spin period. Formally, the age of the nebula is consistent
with the pulsar characteristic age.

{\bf Kes 75.} Basing on the detailed analysis of the distance towards this
object \cite{kes75} conclude that the age of the SNR is in correspondence
with the radio pulsar age.

{\bf G54.1+0.3.} This is a Crab-like nebula. \cite{g54.1+0.3} reported
that no independent reliable age estimate exists for this source.
Roughly, the age of the nebula is consistent with the pulsar spin-down
age.

{\bf CTA 1.}  \cite{cta1} report an age of 13,000 yrs for the SNR
age, referring to \cite{Slane2004}. This value corresponds to a
distance to the remnant of 1.4 kpc. The uncertainty, according to
\cite{Slane2004}, is $\pm 0.3$~kpc, which translates in to $\pm
2800$~yrs for the uncertainty in the age.

{\bf 3C 58.} In a recent paper \cite{3c58} report the value of
5400 yrs referring to earlier studies. \cite{bietenholz2006} gives
7000 yrs with a 3$\sigma$ lower limit of 4300 years.
\cite{rudie2007} suggest a value of ``about few thousand years''.
Finally, \cite{chevalier2005} has $2400\pm 500$ yrs. In our study
we use the interval  4300--7000 yrs.

{\bf S147.} Old age estimates (see, for example, \citealt{kundu1980})
suggested a SNR age of $\sim 80$~--~200~kyrs. In a recent paper
\cite{ng2007} gave arguments in favour of shorter age, $\sim 40$~kyrs. The
kinematic age of the pulsar is also close to this value.

{\bf 0540-693.} Several estimates are available for this source.
\cite{0540-693} give 760$\pm$100 yrs referring to older studies,
in particular to \cite{reynolds1985} and \cite{kirshner1989}.
\cite{reynolds1985} presents an estimate of 800-1100 years.
\cite{kirshner1989} give $762\pm 100$ yrs. \cite{williams2008} and
\cite{park2010} provide wide lists of references and age estimates
for this source. For our study we use an interval 660-1100 yrs.

{\bf Monogem Ring.} \cite{monogem} give a single value, 86,000 yrs, as
the age estimate for this object. They refer to \cite{plucinsky1996},
where this value corresponds to a distance of 300 pc. However, possible
distances range  from 100 to 1300 pc. This can translates in
to a 29,000-371,000 yrs interval for the age estimate. These authors state
that small and large distances are not very probable, but $d=600$~pc is
certainly as good as 300 pc. This distance estimate (600 pc)
corresponds to the age 170,000 yrs.  So, we
use the interval 86-170 kyrs.

{\bf Puppis A.} \cite{puppisa} refer to \cite{wtki1988}, where
an estimate of $3700\pm 300$~yrs is given. We use in our analysis the
range 3400-4100 yrs reported by \cite{puppisa}.

{\bf Vela.}
For the Vela pulsar the upper limit for the age is taken from
\cite{vela1995}. These authors used $d=500$~pc, a value which is
currently revised towards lower values which result in a smaller upper
limit for the age. The lower limit for this SNR
is not very certain, and we use the value 11 kyr. Slightly smaller values
are possible, but this would not change our conclusions about the initial
period significantly. Strictly speaking, the age of the Vela remnant can
be considered just in correspondence with the characteristic age of the
pulsar. Still, we provide a full range for age estimates reported in the
literature, and use the source in our analysis.

{\bf G292.0+1.8.}  \cite{g292.0+1.8} give the range 2400-2850 yrs, which we
use for our estimates.

{\bf G320.4-1.2.} The estimates of the SNR age given by \cite{g320.4-1.2}
are based on standard assumptions about the SN explosion energy  and
ISM density. So, the age estimate is
rather uncertain, probably the nebula's
age is in correspondence with the pulsar age. We do not provide
any estimated of $P_0$ for this object.

{\bf G7.5-1.7.} The estimate of $10^4$~--$10^5$~yrs given by
\cite{g7.5-1.7} is based on numerical estimates for SNRs of this
type, and as such is quite rough. The authors also suggest 50,000 yrs as
a representative value. We use the full range above.

{\bf G12.8-0.0.} We use the range of ages from \cite{brogan2005}.
In \cite{g12.8-0.0} the properties of the radio
pulsar are discussed in details, and arguments in favour of significant
initial period (i.e., compatible with the present day one) are provided.

{\bf G21.5-0.9} \cite{g21.5-0.9} provide a list of
references for age estimates of the SNR. We use the full reported range
which is rather wide.

{\bf W44.} \cite{w44} present two possibilities. The first one gives a
SNR age of 5600-7500 yrs with a central value 6500~yrs, the second
one 19000-25000~yrs. In a recent paper \cite{abdo2010} used an estimate
$\sim$~20000~yrs. For our estimates we use the range 6.5-20~kyrs.

{\bf G65.1+0.6.} The large uncertainty in the age of the SNR
is not influential on our results, as anyway the values are much smaller
than the value of the characteristic age of the pulsar.

{\bf CTB80.} There is only an age estimate for the
pulsar wind nebula. We consider this as a lower limit for the actual age
of the pulsar, so only an upper limit for $P_0$ can be obtained.

{\bf G308.8-0.1.} \cite{g308.8-0.1} report just an
 upper limit for $\tau_{\mathrm{SNR}}$ because the value of the ISM density
is uncertain (in particular it can be lower).
This estimate is larger than
$\tau_{\mathrm{sd}}$, and no conclusion about $P_0$ can be
reached.

{\bf G106.6+2.9.} \cite{g106.6+2.9} obtained an estimate
for the pulsar wind nebula age. We use this as a lower limit for the
age of the pulsar.

{\bf Crab.} The age of this SNR is well-known as it is related to a
historical SN.

{\bf G296.5+10.0.}  \cite{g296.5+10.0} give an estimate of $\sim10^4$~yrs
for an explosion energy $E=2 \, 10^{51}$~erg and an ISM density of
0.2~cm$^{-3}$. They
refer to \cite{kellett1987} who provide an estimate 20,000 yrs for $E=1.5
\, 10^{51}$~erg and ISM density of 0.26~cm$^{-3}$. Both values are
extremely small
in comparison with the spin-down age of the pulsar.

{\bf G315.9-0.0.} \cite{g315.9-0.0} calculate an age estimate
basing on the assumption that the remnant is at the Sedov stage of
expansion. Correspondingly, uncertainties are related to
uncertainties in distance, ISM density, and SN explosion energy,
which are however not discussed.

{\bf G11.2-0.3.} The age of this SNR is considered to be well established as
it is a historical event \citep{g11.2-0.3}. In addition, Sedov phase
estimates are consistent with the historical interpretation
\citep{vasisht1996}.

{\bf Kes 79.} Uncertainties in the age for this SNR are related to
uncertainties in its distance, which is of the order of $\pm 1$~kpc.
Taking into account that the former are at the  $\sim 15-20$\% level
and that the SNR age is orders of magnitude smaller than the characteristic
age of the pulsar, we neglect the uncertainty.

{\bf G78.2+2.1.} Exact
uncertainty in the age is not reported by \cite{g78.2+2.1},
but it is anyway small and does not influence our results.

{\bf G114.3+0.3.}  Exact uncertainty in the age is not reported by
\cite{g114.3+0.3},
but it is anyway small and does not influence our results.

\begin{deluxetable}{lcccc}
\tablewidth{0pt}
\tablecaption{Sample of PSRs associated with SNRs}
\tablehead{
\colhead{PSR} & \colhead{SNR} & \colhead{$\tau_{SNR}/10^3$~yrs} &
\colhead{$\tau_{sd}/10^3$~yrs} & \colhead{Ref.}}
\startdata
J0537-6910 & N157B & as the PSR & 4.9 &  \cite{wg1998}\\
J1119-6127  & G292.2-0.5 & as the PSR & 1.6 & \cite{g292.2-0.5}  \\
J1747-2809 & G0.9+0.1 & as the PSR & 5.3 & \cite{g0.9+0.1} \\
           &          &             &    & \cite{porquet2003}\\
J1747-2958 & G359.23-0.82 & as the PSR & 25.5 & \cite{g359.23-0.82} \\
J1846-0258 & Kes75 & as the PSR & 0.73 & \cite{kes75}\\
J1930+1852 & G54.1+0.3 & as the PSR & 2.9 & \cite{g54.1+0.3}\\
& & & & \\
J0007+7303 & CTA 1 & 10.2-15.8 & 13.9 & \cite{Slane2004}\\
J0205+6449 & 3C58 & 4.3-7 & 5.4 & \cite{3c58}  \\
J0538+2817 & S147 & 40-200 & 618.1 & \cite{s147}\\
           &      &        &       & \cite{ng2007}\\
B0540-69 & 0540-693 & 0.66-1.1 & 1.67 & \cite{williams2008}\\
B0656+14 & Monogem Ring & 86-170 & 110.9 & \cite{monogem}\\
J0821-4300 & Puppis A & 3.3-4.1 & 1489. & \cite{puppisa}\\
B0833-45 & Vela & 11-27 & 11.3 & \cite{vela1995} \\
J1124-5916 & G292.0+1.8 & 2.4-2.85 & 2.85 & \cite{g292.0+1.8}\\
B1509-58  & G320.4-1.2 & 6-20 & 1.6 & \cite{g320.4-1.2}\\
J1809-2332 & G7.5-1.7 & 10-100 & 67.6 & \cite{g7.5-1.7}\\
J1813-1749 & G12.8-0.0 & 0.285-2.5 &  4.7 & \cite{brogan2005}  \\
J1833-1034 & G21.5-0.9 & 0.8-40. & 4.9 & \cite{g21.5-0.9}\\
B1853+01 & W44 & 6.5-20 & 20.3 & \cite{w44} \\
J1957+2831  & G65.1+0.6 & 40-140  & 1568. & \cite{g65.1+0.6}\\
& & & & \\
B1951+32 & CTB80 & $>18$ & 107. & \cite{ctb80} \\
B1338-62 & G308.8-0.1 & $<32.5$ & 12.1 & \cite{g308.8-0.1}\\
J2229+6114 & G106.6+2.9 & $>3.9$ & 10.5 & \cite{g106.6+2.9}\\
& & & & \\
B0531+21 & Crab & 0.957 & 1.24 & \cite{snr2002} \\
J1210-5226 & G296.5+10.0 & 10-20 & 101817. & \cite{g296.5+10.0} \\
J1437-5959 & G315.9-0.0 & 22 & 114. & \cite{g315.9-0.0}\\
J1811-1925 & G11.2-0.3 & 1.6 & 23.2 & \cite{g11.2-0.3}\\
J1852+0040 & Kes79 & 6 & 191502. & \cite{kes79}\\
J2021+4026 & G78.2+2.1 & 6.6 & 76.9 & \cite{g78.2+2.1}\\
B2334+61  & G114.3+0.3 & 7.7 & 40.6 & \cite{g114.3+0.3} \\
\enddata
\end{deluxetable}

\begin{deluxetable}{lcccccc}
\tablecaption{Spin parameters of PSRs in the sample}
\tablehead{
\colhead{PSR} & \colhead{$P$~s} & \colhead{$\dot P$} & \colhead{$B/10^{12}$~G}
& \colhead{$P_0$~s} & \colhead{$P_0/P$}}
\startdata
J0537-6910 & 0.016 & 5.18E-14 &  0.92& $\ll P$&  $\sim 0$\\
J1119-6127 & 0.408 & 4.02E-12 & 41.    &  $\ll P$ &  $\sim 0$\\
J1747-2809 &  0.052 & 1.56E-13 & 2.9 & $\ll P$&  $\sim 0$\\
J1747-2958 & 0.099 & 6.13E-14 &   2.5 &  $\ll P$ &  $\sim 0$ \\
J1846-0258 & 0.326 & 7.08E-12 & 48.6 & $\ll P$&  $\sim 0$\\
J1930+1852 & 0.137 & 7.51E-13 & 10.3 &  $\ll P$ &  $\sim 0$ \\
& & & & &  \\
J0007+7303 & 0.316 & 3.6E-13& 10.8 &  $<0.163$& $<0.52$ \\
J0205+6449 & 0.066 & 1.94E-13& 3.6 & $<0.029$& $<0.45$ \\
J0538+2817 & 0.143 & 3.67E-15 & 0.73 & $<0.134$ &  $<0.93$\\
  & 0.143 & 3.67E-15 & 0.73 & $>0.118$ &  $>0.82$\\
B0540-69  & 0.05 & 4.79E-13 & 5.0 & $<0.039$  & $<0.78$\\
& 0.05& 4.79E-13& 5.0& $>0.03$& $>0.59$ \\
B0656+14 & 0.385 & 5.5E-14  & 4.7 & $<0.183$  & $<0.48$\\
J0821-4300 & 0.113 &  1.2E-15 & 0.37 & $<0.113$ & $\sim 1$\\
& 0.113 &  1.2E-15 & 0.37 & $>0.113$ &  $\sim 1$\\
B0833-45 &   0.089 & 1.25E-13 & 3.4 & $<0.016$ &  $<0.2$\\
J1124-5916 & 0.135 & 7.53E-13 & 10.2 & $<0.054$  & $<0.40$ \\
& 0.135&7.53E-13 &10.2 & $>0.004$& $>0.03$ \\
J1210-5226 & 0.424 & 6.6E-17 & 0.17 & 0.424  & $\sim 1$\\
B1509-58 & 0.151 & 1.54E-12 & 15.4 & ---&  ---\\
J1809-2332 &  0.147 & 3.44E-14 & 2.3 & $<0.136$ &  $<0.92$\\
J1813-1749 & 0.045 & 1.5E-13 & 2.6 & $<0.043$ &  $<0.97$ \\
& 0.045 & 1.5E-13 & 2.6 & $>0.031$ &  $>0.69$\\
J1833-1034  & 0.062 & 2.02E-13 & 3.6 & $<0.057$ &  $<0.91$\\
B1853+01 &   0.267 & 2.08E-13 & 7.5 & $<0.221$ &  $<0.83$\\
& 0.267 & 2.08E-13 & 7.5 & $>0.036$ &  $>0.14$\\
J1957+2831  & 0.308 & 3.11E-15 & 0.99 & $<0.3$ &  $<0.99$\\
&  0.308 & 3.11E-15 & 0.99 &  $>0.29$ &  $>0.95$\\
& & & & &  \\
B1951+32 &   0.04  & 5.84E-15 & 0.49 & $<0.036$  & $<0.91$ \\
B1338-62 & 0.193 & 2.53E-13 & 7.1 & ---  & --- \\
J2229+6114 & 0.052 & 7.83E-14 & 2.0 & $<0.041$&   $<0.79$\\
& & & & & \\
B0531+21 &   0.033 & 4.23E-13 &  3.8 & 0.016  & 0.48 \\
J1437-5959 & 0.062 & 8.59E-15 & 0.74 & 0.055  & 0.9\\
J1811-1925 & 0.065 & 4.40E-14  & 1.7 & 0.062  & 0.97\\
J1852+0040  & 0.105 & 8.68E-18  & 0.03 & 0.105  & $\sim 1$\\
J2021+4026 & 0.265 & 5.47E-14 & 3.9 & 0.254  & 0.96\\
B2334+61  & 0.495 & 1.93E-13 & 9.9 & 0.45  & 0.91 \\
\enddata
\end{deluxetable}

\section{Discussion}


For our estimates we use a braking index $n=3$. Additional checks
were made for other values, $2<n< 10$, and we found that the
derived values of $P_0$ are not significantly changed in many
cases. For $n=2$ $P_0$ may be larger by a factor up to 2, but
these changes do not modify the general picture shown in Fig. 1
significantly.
This is an expected result, since the objects under
consideration are very young, and is in correspondence with the
conclusions by \cite{fgkaspi} that varying $n$ around the standard
value has a weak impact on the parameters of the pulsar
population. However, with growing $n$  more and more SNR ages are
found to be in contradiction with the  spin-down ages. Already
for $n=4$ in several cases $\tau_\mathrm{sd}$ become smaller than
$\tau_\mathrm{SNR}$.


Recently,  an e-print by \cite{china} appeared in which a
correlation between magnetic field and the ratio of the ``true''
(i.e. $\tau_\mathrm{SNR}$)  and spin-down ages was proposed. The authors
explain this correlation as due to field decay. We present a similar
plot, i.e. $B$ vs. ages ratio, in Fig. 2 using data from Table 2.
However, we believe that the
existence of such a correlation is an artifact, and the reason for
its appearance is that \cite{china}  neglect the
initial spin period.
This conclusion can be illustrated by plotting $B$ vs. the difference
$P-P_0$.  Clearly, in the sample of young objects (associated with SNRs)
those with smaller magnetic fields
have initial spin periods closer to the present day periods.

Naturally, objects with lower field, like the ``anti-magnetars''
associated to some central compact objects in SNRs, have present spin periods
very close to the initial values. Then, the usual estimate of the
spin-down age assuming $P_0\simeq 0$ is invalid, as the true age
is very different.

\begin{figure}
\epsscale{.99}
\plotone{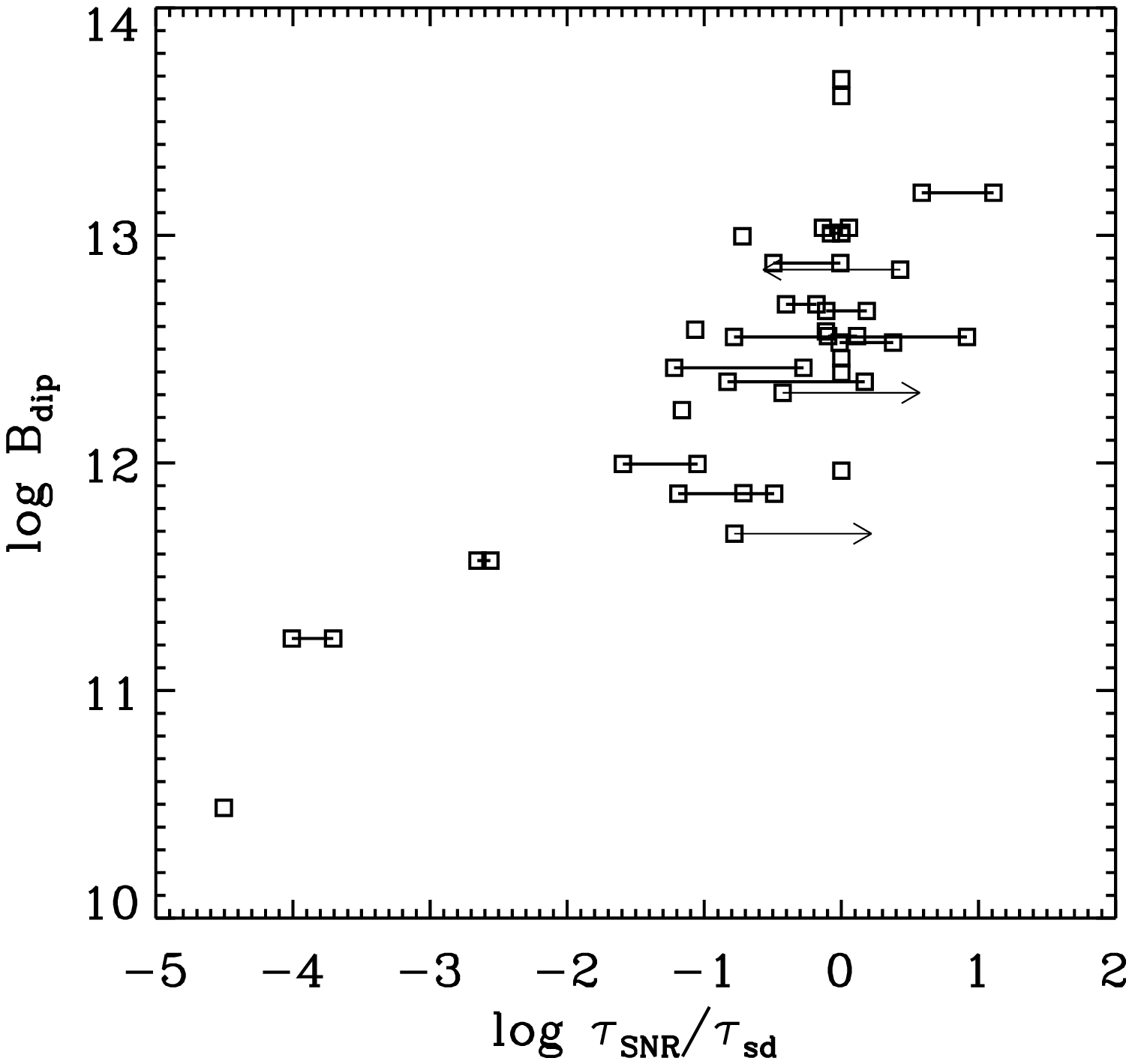}
\caption{A plot similar to the one in \cite{china}: magnetic field vs.
the ratio of ``true'' age ($\tau_\mathrm{SNR}$) and spin-down age.
\label{fig2}}
\end{figure}




We cannot exclude that there was some field evolution in the
objects under discussion (note that field evolution is
undistinguishable from the evolution of the angle between spin and
magnetic axis), but we see no evidence for it, contrary to the conclusion by
\cite{china}.

Not surprisingly, the assumptions made by \cite{china} led them to
suggest \citep{xie2012} that field decay is so strong and
important that in their youth most of standard pulsars have magnetar
properties, $B\sim 5 \, 10^{14}$~G at the age $\sim 0.4$~yrs. We
believe that this is not needed if non-zero initial
spin periods are accounted for.

Oscillations of braking indices were also proposed by
\cite{biryukov2012} on a time scale $\sim 10^3$~--~$10^4$~yrs.
If this is valid also for young pulsars, then our estimates of
$P_0$ must be corrected. But still, non-zero $P_0$ will remain the
main reason to explain discrepancy between spin-down ages and SNR
ages.


In Fig. 1 we see that under the assumptions we made no correlation between
$P_0$ and $B$ is visible.  This is in correspondence with standard
assumptions made in different kinds of models.
We did not try to figure out  a possible shape of the
initial spin period distribution (for example, if it is a multi-peak
distribution or not --- in the future, probably, it will appear that, for
example, so-called ``anti-magnetars'' form a separate subpopulation, and so
they do not fit the same single-mode distribution as normal radio pulsars;
if gaussian is better than log-gaussian, symmetric
distributions better than asymmetric, etc.).
However, the data we obtained
can be used to check assumptions about initial spin distributions.
For example, obviously very narrow distributions are not in
correspondence with the data as there are pulsars with $P_0$ from
milliseconds to hundreds of milliseconds.

In order to derive information on the distribution of the initial periods
from
the values derived from observations, we run a number of Kolmogorov-Smirnov
(KS) tests, using different forms of the distribution function. In
particular,
we tested gaussian distributions with different parameters and a top-hat
(i.e. constant in a range and zero outside) distribution.
For this calculations we used data on 11 PSRs.
Six of the are those with an exactly determined $P_0$, to which we
added five objects with small uncertainty for the initial period value using
average SNR age.

\begin{figure}
\epsscale{.99}
\plotone{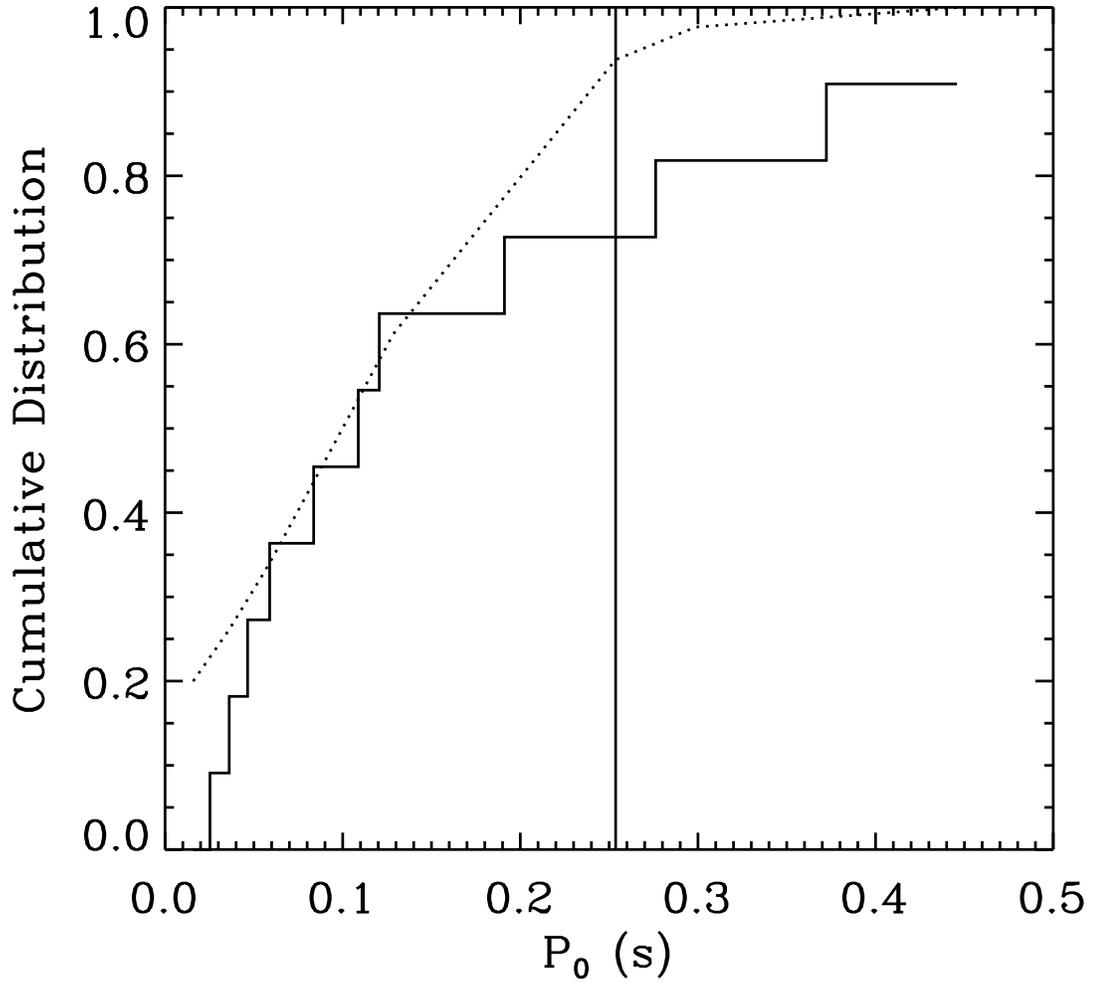}
\caption{The data cumulative distribution function (full histogram) compared
with that
of a gaussian distribution with average 0.1 s and standard deviation of 0.1
s
(dotted line). The vertical line shows the value of $P_0$ where the
difference
is largest.
\label{fig3}}
\end{figure}

\begin{figure}
\epsscale{.99}
\plotone{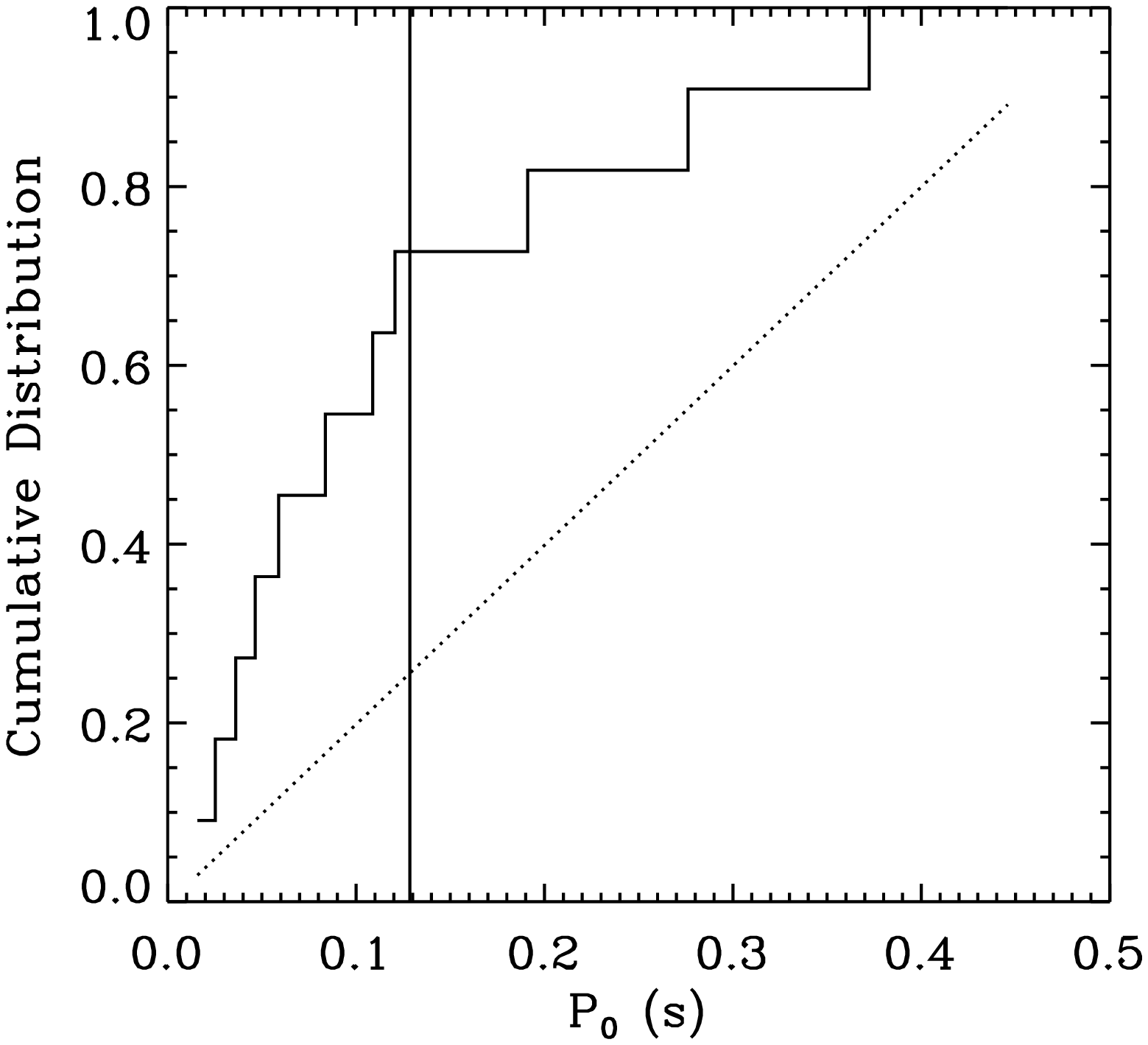}
\caption{Same as in Fig. \ref{fig3} for  a top-hat distribution in the range
$0.001\, {\rm
s}\leq P_0\leq 0.5\, {\rm s}$.
\label{fig4}}
\end{figure}

Two representative
examples are shown in Figs. \ref{fig3}--\ref{fig4}.
 The gaussian distribution of  Fig. \ref{fig3}
 gives a
KS significance level of $\sim 0.66$, indicating that the data and model
have
quite consistent cumulative distributions. On the other hand, the top-hat
distribution has a very low value of the KS significance, $< 0.01$, and
should
be rejected.

\section{Conclusions}

In this paper we discussed a sample of radio pulsars associated with
supernova remnants. This association allows to obtain estimates of initial
spin periods in $\ga$~20 cases. The obtained distribution is rather flat in
wide range from $\sim 10$~msec to hundreds of msec.

\acknowledgments
We thank Andrea Possenti for some helpful comments on the
paper. SP also thanks Anton Biryukov and Andrei Igoshev for discussions.
The work of S.P. was supported by RFBR grants (10-02-00599, 12-02-00186)
and by the Federal programm for scientific and educational staff
(02.740.11.0575). RT acknowledges financial support by INAF through a
PRIN 2011 grant. SP thanks the University of Padova for hospitality.

\bibliography{Popov_bib}

\begin{thebibliography}{61}
\ifx \bisbn   \undefined \def \bisbn  #1{ISBN #1}\fi
\ifx \binits  \undefined \def \binits#1{#1} \fi
\ifx \bauthor  \undefined \def \bauthor#1{#1} \fi
\ifx \batitle  \undefined \def \batitle#1{#1} \fi
\ifx \bjtitle  \undefined \def \bjtitle#1{#1}\fi
\ifx \bvolume  \undefined \def \bvolume#1{\textbf{#1}}\fi
\ifx \byear  \undefined \def \byear#1{#1} \fi
\ifx \bissue  \undefined \def \bissue#1{#1} \fi
\ifx \bfpage  \undefined \def \bfpage#1{#1} \fi
\ifx \blpage  \undefined \def \blpage #1{#1} \fi
\ifx \burl  \undefined \def \burl#1{\textsf{#1}} \fi
\ifx \doiurl  \undefined \def \doiurl#1{\textsf{#1}} \fi
\ifx \betal  \undefined \def \betal{\textit{et al.}} \fi
\ifx \binstitute  \undefined \def \binstitute#1{#1} \fi
\ifx \binstitutionaled  \undefined \def \binstitutionaled#1{#1} \fi
\ifx \bctitle  \undefined \def \bctitle#1{#1} \fi
\ifx \beditor  \undefined \def \beditor#1{#1} \fi
\ifx \bpublisher  \undefined \def \bpublisher#1{#1} \fi
\ifx \bbtitle  \undefined \def \bbtitle#1{#1} \fi
\ifx \bedition  \undefined \def \bedition#1{#1} \fi
\ifx \bseriesno  \undefined \def \bseriesno#1{#1} \fi
\ifx \blocation  \undefined \def \blocation#1{#1} \fi
\ifx \bsertitle  \undefined \def \bsertitle#1{#1} \fi
\ifx \bsnm \undefined \def \bsnm#1{#1} \fi
\ifx \bsuffix \undefined \def \bsuffix#1{#1} \fi
\ifx \bparticle \undefined \def \bparticle#1{#1} \fi
\ifx \barticle \undefined \def \barticle#1{#1} \fi
\ifx \bconfdate \undefined \def \bconfdate #1{#1} \fi
\ifx \botherref \undefined \def \botherref #1{#1} \fi
\ifx \url \undefined \def \url#1{\textsf{#1}} \fi
\ifx \bchapter \undefined \def \bchapter#1{#1} \fi
\ifx \bbook \undefined \def \bbook#1{#1} \fi
\ifx \bcomment \undefined \def \bcomment#1{#1} \fi
\ifx \oauthor \undefined \def \oauthor#1{#1} \fi
\ifx \citeauthoryear \undefined \def \citeauthoryear#1{#1} \fi
\ifx \endbibitem  \undefined \def \endbibitem {}\fi
\ifx \bconflocation  \undefined \def \bconflocation#1{#1} \fi
\ifx \arxivurl  \undefined \def \arxivurl#1{\textsf{#1}} \fi

\bibitem[\protect\citeauthoryear{{Abdo} and {et al.}}{2010}]{abdo2010}
\begin{barticle}
\bauthor{\bsnm{{Abdo}}, \binits{A.A.}}, \betal:
\bjtitle{Science}
\bvolume{327},
\bfpage{1103}
(\byear{2010}).
doi:\doiurl{10.1126/science.1182787}
\end{barticle}
\endbibitem

\bibitem[\protect\citeauthoryear{{Aharonian} and {et al.}}{2005}]{g0.9+0.1}
\begin{barticle}
\bauthor{\bsnm{{Aharonian}}, \binits{F.}}, \betal:
\bjtitle{\aap}
\bvolume{432},
\bfpage{25}
(\byear{2005}).
\arxivurl{arXiv:astro-ph/0501265}.
doi:\doiurl{10.1051/0004-6361:200500022}
\end{barticle}
\endbibitem

\bibitem[\protect\citeauthoryear{{Anderson} et~al.}{1996}]{s147}
\begin{barticle}
\bauthor{\bsnm{{Anderson}}, \binits{S.B.}},
\bauthor{\bsnm{{Cadwell}}, \binits{B.J.}},
\bauthor{\bsnm{{Jacoby}}, \binits{B.A.}},
\bauthor{\bsnm{{Wolszczan}}, \binits{A.}},
\bauthor{\bsnm{{Foster}}, \binits{R.S.}},
\bauthor{\bsnm{{Kramer}}, \binits{M.}}:
\bjtitle{\apjl}
\bvolume{468},
\bfpage{55}
(\byear{1996}).
doi:\doiurl{10.1086/310218}
\end{barticle}
\endbibitem

\bibitem[\protect\citeauthoryear{{Aschenbach} et~al.}{1995}]{vela1995}
\begin{barticle}
\bauthor{\bsnm{{Aschenbach}}, \binits{B.}},
\bauthor{\bsnm{{Egger}}, \binits{R.}},
\bauthor{\bsnm{{Tr{\"u}mper}}, \binits{J.}}:
\bjtitle{\nat}
\bvolume{373},
\bfpage{587}
(\byear{1995}).
doi:\doiurl{10.1038/373587a0}
\end{barticle}
\endbibitem

\bibitem[\protect\citeauthoryear{{Bietenholz}}{2006}]{bietenholz2006}
\begin{barticle}
\bauthor{\bsnm{{Bietenholz}}, \binits{M.F.}}:
\bjtitle{\apj}
\bvolume{645},
\bfpage{1180}
(\byear{2006}).
\arxivurl{arXiv:astro-ph/0603197}.
doi:\doiurl{10.1086/504584}
\end{barticle}
\endbibitem

\bibitem[\protect\citeauthoryear{{Biryukov} et~al.}{2012}]{biryukov2012}
\begin{barticle}
\bauthor{\bsnm{{Biryukov}}, \binits{A.}},
\bauthor{\bsnm{{Beskin}}, \binits{G.}},
\bauthor{\bsnm{{Karpov}}, \binits{S.}}:
\bjtitle{\mnras}
\bvolume{420},
\bfpage{103}
(\byear{2012}).
\arxivurl{1105.5019}.
doi:\doiurl{10.1111/j.1365-2966.2011.20005.x}
\end{barticle}
\endbibitem

\bibitem[\protect\citeauthoryear{{Brogan} et~al.}{2005}]{brogan2005}
\begin{barticle}
\bauthor{\bsnm{{Brogan}}, \binits{C.L.}},
\bauthor{\bsnm{{Gaensler}}, \binits{B.M.}},
\bauthor{\bsnm{{Gelfand}}, \binits{J.D.}},
\bauthor{\bsnm{{Lazendic}}, \binits{J.S.}},
\bauthor{\bsnm{{Lazio}}, \binits{T.J.W.}},
\bauthor{\bsnm{{Kassim}}, \binits{N.E.}},
\bauthor{\bsnm{{McClure-Griffiths}}, \binits{N.M.}}:
\bjtitle{\apjl}
\bvolume{629},
\bfpage{105}
(\byear{2005}).
\arxivurl{arXiv:astro-ph/0505145}.
doi:\doiurl{10.1086/491471}
\end{barticle}
\endbibitem

\bibitem[\protect\citeauthoryear{{Camilo} et~al.}{2002a}]{g54.1+0.3}
\begin{barticle}
\bauthor{\bsnm{{Camilo}}, \binits{F.}},
\bauthor{\bsnm{{Lorimer}}, \binits{D.R.}},
\bauthor{\bsnm{{Bhat}}, \binits{N.D.R.}},
\bauthor{\bsnm{{Gotthelf}}, \binits{E.V.}},
\bauthor{\bsnm{{Halpern}}, \binits{J.P.}},
\bauthor{\bsnm{{Wang}}, \binits{Q.D.}},
\bauthor{\bsnm{{Lu}}, \binits{F.J.}},
\bauthor{\bsnm{{Mirabal}}, \binits{N.}}:
\bjtitle{\apjl}
\bvolume{574},
\bfpage{71}
(\byear{2002}a).
\arxivurl{arXiv:astro-ph/0206220}.
doi:\doiurl{10.1086/342351}
\end{barticle}
\endbibitem

\bibitem[\protect\citeauthoryear{{Camilo} et~al.}{2002b}]{g359.23-0.82}
\begin{barticle}
\bauthor{\bsnm{{Camilo}}, \binits{F.}},
\bauthor{\bsnm{{Manchester}}, \binits{R.N.}},
\bauthor{\bsnm{{Gaensler}}, \binits{B.M.}},
\bauthor{\bsnm{{Lorimer}}, \binits{D.R.}}:
\bjtitle{\apjl}
\bvolume{579},
\bfpage{25}
(\byear{2002}b).
\arxivurl{arXiv:astro-ph/0209480}.
doi:\doiurl{10.1086/344832}
\end{barticle}
\endbibitem

\bibitem[\protect\citeauthoryear{{Camilo} et~al.}{2009}]{g315.9-0.0}
\begin{barticle}
\bauthor{\bsnm{{Camilo}}, \binits{F.}},
\bauthor{\bsnm{{Ng}}, \binits{C.-Y.}},
\bauthor{\bsnm{{Gaensler}}, \binits{B.M.}},
\bauthor{\bsnm{{Ransom}}, \binits{S.M.}},
\bauthor{\bsnm{{Chatterjee}}, \binits{S.}},
\bauthor{\bsnm{{Reynolds}}, \binits{J.}},
\bauthor{\bsnm{{Sarkissian}}, \binits{J.}}:
\bjtitle{\apjl}
\bvolume{703},
\bfpage{55}
(\byear{2009}).
\arxivurl{0908.2421}.
doi:\doiurl{10.1088/0004-637X/703/1/L55}
\end{barticle}
\endbibitem

\bibitem[\protect\citeauthoryear{{Castelletti} et~al.}{2003}]{ctb80}
\begin{barticle}
\bauthor{\bsnm{{Castelletti}}, \binits{G.}},
\bauthor{\bsnm{{Dubner}}, \binits{G.}},
\bauthor{\bsnm{{Golap}}, \binits{K.}},
\bauthor{\bsnm{{Goss}}, \binits{W.M.}},
\bauthor{\bsnm{{Vel{\'a}zquez}}, \binits{P.F.}},
\bauthor{\bsnm{{Holdaway}}, \binits{M.}},
\bauthor{\bsnm{{Rao}}, \binits{A.P.}}:
\bjtitle{\aj}
\bvolume{126},
\bfpage{2114}
(\byear{2003}).
\arxivurl{arXiv:astro-ph/0310655}.
doi:\doiurl{10.1086/378482}
\end{barticle}
\endbibitem

\bibitem[\protect\citeauthoryear{{Caswell} et~al.}{1992}]{g308.8-0.1}
\begin{barticle}
\bauthor{\bsnm{{Caswell}}, \binits{J.L.}},
\bauthor{\bsnm{{Kesteven}}, \binits{M.J.}},
\bauthor{\bsnm{{Stewart}}, \binits{R.T.}},
\bauthor{\bsnm{{Milne}}, \binits{D.K.}},
\bauthor{\bsnm{{Haynes}}, \binits{R.F.}}:
\bjtitle{\apjl}
\bvolume{399},
\bfpage{151}
(\byear{1992}).
doi:\doiurl{10.1086/186629}
\end{barticle}
\endbibitem

\bibitem[\protect\citeauthoryear{{Chevalier}}{2005}]{chevalier2005}
\begin{barticle}
\bauthor{\bsnm{{Chevalier}}, \binits{R.A.}}:
\bjtitle{\apj}
\bvolume{619},
\bfpage{839}
(\byear{2005}).
\arxivurl{arXiv:astro-ph/0409013}.
doi:\doiurl{10.1086/426584}
\end{barticle}
\endbibitem

\bibitem[\protect\citeauthoryear{{Dean} and {Hill}}{2008}]{g12.8-0.0}
\begin{barticle}
\bauthor{\bsnm{{Dean}}, \binits{A.J.}},
\bauthor{\bsnm{{Hill}}, \binits{A.B.}}:
\bjtitle{\aap}
\bvolume{485},
\bfpage{195}
(\byear{2008}).
\arxivurl{0804.3420}.
doi:\doiurl{10.1051/0004-6361:200809356}
\end{barticle}
\endbibitem

\bibitem[\protect\citeauthoryear{{Faucher-Gigu{\`e}re} and
  {Kaspi}}{2006}]{fgkaspi}
\begin{barticle}
\bauthor{\bsnm{{Faucher-Gigu{\`e}re}}, \binits{C.-A.}},
\bauthor{\bsnm{{Kaspi}}, \binits{V.M.}}:
\bjtitle{\apj}
\bvolume{643},
\bfpage{332}
(\byear{2006}).
\arxivurl{arXiv:astro-ph/0512585}.
doi:\doiurl{10.1086/501516}
\end{barticle}
\endbibitem

\bibitem[\protect\citeauthoryear{{Fryer} et~al.}{2011}]{fryer2011}
\begin{botherref}
\oauthor{\bsnm{{Fryer}}, \binits{C.L.}},
\oauthor{\bsnm{{Belczynski}}, \binits{K.}},
\oauthor{\bsnm{{Wiktorowicz}}, \binits{G.}},
\oauthor{\bsnm{{Dominik}}, \binits{M.}},
\oauthor{\bsnm{{Kalogera}}, \binits{V.}},
\oauthor{\bsnm{{Holz}}, \binits{D.}}:
ArXiv e-prints
(2011).
\arxivurl{1110.1726}
\end{botherref}
\endbibitem

\bibitem[\protect\citeauthoryear{{Gonzalez} and {Safi-Harb}}{2003}]{g292.0+1.8}
\begin{barticle}
\bauthor{\bsnm{{Gonzalez}}, \binits{M.}},
\bauthor{\bsnm{{Safi-Harb}}, \binits{S.}}:
\bjtitle{\apjl}
\bvolume{583},
\bfpage{91}
(\byear{2003}).
\arxivurl{arXiv:astro-ph/0301193}.
doi:\doiurl{10.1086/368122}
\end{barticle}
\endbibitem

\bibitem[\protect\citeauthoryear{{Gotthelf} and {Halpern}}{2009}]{puppisa}
\begin{barticle}
\bauthor{\bsnm{{Gotthelf}}, \binits{E.V.}},
\bauthor{\bsnm{{Halpern}}, \binits{J.P.}}:
\bjtitle{\apjl}
\bvolume{695},
\bfpage{35}
(\byear{2009}).
\arxivurl{0902.3007}.
doi:\doiurl{10.1088/0004-637X/695/1/L35}
\end{barticle}
\endbibitem

\bibitem[\protect\citeauthoryear{{Green} and {Stephenson}}{2003}]{green2003}
\begin{bchapter}
\bauthor{\bsnm{{Green}}, \binits{D.A.}},
\bauthor{\bsnm{{Stephenson}}, \binits{F.R.}}:
In: \beditor{\bsnm{{K.~Weiler}}} (ed.)
\bbtitle{Supernovae and Gamma-Ray Bursters}.
\bsertitle{Lecture Notes in Physics, Berlin Springer Verlag},
vol. \bseriesno{598},
p. \bfpage{7}
(\byear{2003}).
\arxivurl{arXiv:astro-ph/0301603}
\end{bchapter}
\endbibitem

\bibitem[\protect\citeauthoryear{{Haensel} et~al.}{2007}]{hyp2007}
\begin{bbook}
\beditor{\bsnm{{Haensel}}, \binits{P.}},
\beditor{\bsnm{{Potekhin}}, \binits{A.Y.}},
\beditor{\bsnm{{Yakovlev}}, \binits{D.G.}} (eds.):
\bbtitle{{Neutron Stars 1 : Equation of State and Structure}}.
\bsertitle{Astrophysics and Space Science Library},
vol. \bseriesno{326}
(\byear{2007})
\end{bbook}
\endbibitem

\bibitem[\protect\citeauthoryear{{Halpern} et~al.}{2004}]{cta1}
\begin{barticle}
\bauthor{\bsnm{{Halpern}}, \binits{J.P.}},
\bauthor{\bsnm{{Gotthelf}}, \binits{E.V.}},
\bauthor{\bsnm{{Camilo}}, \binits{F.}},
\bauthor{\bsnm{{Helfand}}, \binits{D.J.}},
\bauthor{\bsnm{{Ransom}}, \binits{S.M.}}:
\bjtitle{\apj}
\bvolume{612},
\bfpage{398}
(\byear{2004}).
\arxivurl{arXiv:astro-ph/0404312}.
doi:\doiurl{10.1086/422409}
\end{barticle}
\endbibitem

\bibitem[\protect\citeauthoryear{{Harrus} et~al.}{1997}]{w44}
\begin{barticle}
\bauthor{\bsnm{{Harrus}}, \binits{I.M.}},
\bauthor{\bsnm{{Hughes}}, \binits{J.P.}},
\bauthor{\bsnm{{Singh}}, \binits{K.P.}},
\bauthor{\bsnm{{Koyama}}, \binits{K.}},
\bauthor{\bsnm{{Asaoka}}, \binits{I.}}:
\bjtitle{\apj}
\bvolume{488},
\bfpage{781}
(\byear{1997}).
\arxivurl{arXiv:astro-ph/9705239}.
doi:\doiurl{10.1086/304717}
\end{barticle}
\endbibitem

\bibitem[\protect\citeauthoryear{{Kellett} et~al.}{1987}]{kellett1987}
\begin{barticle}
\bauthor{\bsnm{{Kellett}}, \binits{B.J.}},
\bauthor{\bsnm{{Branduardi-Raymont}}, \binits{G.}},
\bauthor{\bsnm{{Culhane}}, \binits{J.L.}},
\bauthor{\bsnm{{Mason}}, \binits{I.M.}},
\bauthor{\bsnm{{Mason}}, \binits{K.O.}},
\bauthor{\bsnm{{Whitehouse}}, \binits{D.R.}}:
\bjtitle{\mnras}
\bvolume{225},
\bfpage{199}
(\byear{1987})
\end{barticle}
\endbibitem

\bibitem[\protect\citeauthoryear{{Kirshner} et~al.}{1989}]{kirshner1989}
\begin{barticle}
\bauthor{\bsnm{{Kirshner}}, \binits{R.P.}},
\bauthor{\bsnm{{Morse}}, \binits{J.A.}},
\bauthor{\bsnm{{Winkler}}, \binits{P.F.}},
\bauthor{\bsnm{{Blair}}, \binits{W.P.}}:
\bjtitle{\apj}
\bvolume{342},
\bfpage{260}
(\byear{1989}).
doi:\doiurl{10.1086/167590}
\end{barticle}
\endbibitem

\bibitem[\protect\citeauthoryear{{Kothes} et~al.}{2006}]{g106.6+2.9}
\begin{barticle}
\bauthor{\bsnm{{Kothes}}, \binits{R.}},
\bauthor{\bsnm{{Reich}}, \binits{W.}},
\bauthor{\bsnm{{Uyan{\i}ker}}, \binits{B.}}:
\bjtitle{\apj}
\bvolume{638},
\bfpage{225}
(\byear{2006}).
doi:\doiurl{10.1086/498666}
\end{barticle}
\endbibitem

\bibitem[\protect\citeauthoryear{{Kundu} et~al.}{1980}]{kundu1980}
\begin{barticle}
\bauthor{\bsnm{{Kundu}}, \binits{M.R.}},
\bauthor{\bsnm{{Angerhofer}}, \binits{P.E.}},
\bauthor{\bsnm{{Fuerst}}, \binits{E.}},
\bauthor{\bsnm{{Hirth}}, \binits{W.}}:
\bjtitle{\aap}
\bvolume{92},
\bfpage{225}
(\byear{1980})
\end{barticle}
\endbibitem

\bibitem[\protect\citeauthoryear{{Lattimer} and {Prakash}}{2004}]{lp2004}
\begin{barticle}
\bauthor{\bsnm{{Lattimer}}, \binits{J.M.}},
\bauthor{\bsnm{{Prakash}}, \binits{M.}}:
\bjtitle{Science}
\bvolume{304},
\bfpage{536}
(\byear{2004}).
\arxivurl{arXiv:astro-ph/0405262}.
doi:\doiurl{10.1126/science.1090720}
\end{barticle}
\endbibitem

\bibitem[\protect\citeauthoryear{{Leahy} and {Tian}}{2008}]{kes75}
\begin{barticle}
\bauthor{\bsnm{{Leahy}}, \binits{D.A.}},
\bauthor{\bsnm{{Tian}}, \binits{W.W.}}:
\bjtitle{\aap}
\bvolume{480},
\bfpage{25}
(\byear{2008}).
\arxivurl{0711.4107}.
doi:\doiurl{10.1051/0004-6361:20079149}
\end{barticle}
\endbibitem

\bibitem[\protect\citeauthoryear{{Manchester} et~al.}{1993}]{0540-693}
\begin{barticle}
\bauthor{\bsnm{{Manchester}}, \binits{R.N.}},
\bauthor{\bsnm{{Staveley-Smith}}, \binits{L.}},
\bauthor{\bsnm{{Kesteven}}, \binits{M.J.}}:
\bjtitle{\apj}
\bvolume{411},
\bfpage{756}
(\byear{1993}).
doi:\doiurl{10.1086/172877}
\end{barticle}
\endbibitem

\bibitem[\protect\citeauthoryear{{Manchester} et~al.}{2005}]{atnf}
\begin{barticle}
\bauthor{\bsnm{{Manchester}}, \binits{R.N.}},
\bauthor{\bsnm{{Hobbs}}, \binits{G.B.}},
\bauthor{\bsnm{{Teoh}}, \binits{A.}},
\bauthor{\bsnm{{Hobbs}}, \binits{M.}}:
\bjtitle{\aj}
\bvolume{129},
\bfpage{1993}
(\byear{2005}).
doi:\doiurl{10.1086/428488}
\end{barticle}
\endbibitem

\bibitem[\protect\citeauthoryear{{Marshall} et~al.}{1998}]{n157b}
\begin{barticle}
\bauthor{\bsnm{{Marshall}}, \binits{F.E.}},
\bauthor{\bsnm{{Gotthelf}}, \binits{E.V.}},
\bauthor{\bsnm{{Zhang}}, \binits{W.}},
\bauthor{\bsnm{{Middleditch}}, \binits{J.}},
\bauthor{\bsnm{{Wang}}, \binits{Q.D.}}:
\bjtitle{\apjl}
\bvolume{499},
\bfpage{179}
(\byear{1998}).
\arxivurl{arXiv:astro-ph/9803214}.
doi:\doiurl{10.1086/311381}
\end{barticle}
\endbibitem

\bibitem[\protect\citeauthoryear{{Migliazzo} et~al.}{2002}]{migl2002}
\begin{barticle}
\bauthor{\bsnm{{Migliazzo}}, \binits{J.M.}},
\bauthor{\bsnm{{Gaensler}}, \binits{B.M.}},
\bauthor{\bsnm{{Backer}}, \binits{D.C.}},
\bauthor{\bsnm{{Stappers}}, \binits{B.W.}},
\bauthor{\bsnm{{van der Swaluw}}, \binits{E.}},
\bauthor{\bsnm{{Strom}}, \binits{R.G.}}:
\bjtitle{\apjl}
\bvolume{567},
\bfpage{141}
(\byear{2002}).
\arxivurl{arXiv:astro-ph/0202063}.
doi:\doiurl{10.1086/340002}
\end{barticle}
\endbibitem

\bibitem[\protect\citeauthoryear{{Ng} et~al.}{2007}]{ng2007}
\begin{barticle}
\bauthor{\bsnm{{Ng}}, \binits{C.-Y.}},
\bauthor{\bsnm{{Romani}}, \binits{R.W.}},
\bauthor{\bsnm{{Brisken}}, \binits{W.F.}},
\bauthor{\bsnm{{Chatterjee}}, \binits{S.}},
\bauthor{\bsnm{{Kramer}}, \binits{M.}}:
\bjtitle{\apj}
\bvolume{654},
\bfpage{487}
(\byear{2007}).
\arxivurl{arXiv:astro-ph/0611068}.
doi:\doiurl{10.1086/510576}
\end{barticle}
\endbibitem

\bibitem[\protect\citeauthoryear{{Park} et~al.}{2010}]{park2010}
\begin{barticle}
\bauthor{\bsnm{{Park}}, \binits{S.}},
\bauthor{\bsnm{{Hughes}}, \binits{J.P.}},
\bauthor{\bsnm{{Slane}}, \binits{P.O.}},
\bauthor{\bsnm{{Mori}}, \binits{K.}},
\bauthor{\bsnm{{Burrows}}, \binits{D.N.}}:
\bjtitle{\apj}
\bvolume{710},
\bfpage{948}
(\byear{2010}).
\arxivurl{0912.5177}.
doi:\doiurl{10.1088/0004-637X/710/2/948}
\end{barticle}
\endbibitem

\bibitem[\protect\citeauthoryear{{Pivovaroff} et~al.}{2001}]{g292.2-0.5}
\begin{barticle}
\bauthor{\bsnm{{Pivovaroff}}, \binits{M.J.}},
\bauthor{\bsnm{{Kaspi}}, \binits{V.M.}},
\bauthor{\bsnm{{Camilo}}, \binits{F.}},
\bauthor{\bsnm{{Gaensler}}, \binits{B.M.}},
\bauthor{\bsnm{{Crawford}}, \binits{F.}}:
\bjtitle{\apj}
\bvolume{554},
\bfpage{161}
(\byear{2001}).
\arxivurl{arXiv:astro-ph/0102084}.
doi:\doiurl{10.1086/321340}
\end{barticle}
\endbibitem

\bibitem[\protect\citeauthoryear{{Plucinsky} et~al.}{1996}]{plucinsky1996}
\begin{barticle}
\bauthor{\bsnm{{Plucinsky}}, \binits{P.P.}},
\bauthor{\bsnm{{Snowden}}, \binits{S.L.}},
\bauthor{\bsnm{{Aschenbach}}, \binits{B.}},
\bauthor{\bsnm{{Egger}}, \binits{R.}},
\bauthor{\bsnm{{Edgar}}, \binits{R.J.}},
\bauthor{\bsnm{{McCammon}}, \binits{D.}}:
\bjtitle{\apj}
\bvolume{463},
\bfpage{224}
(\byear{1996}).
doi:\doiurl{10.1086/177236}
\end{barticle}
\endbibitem

\bibitem[\protect\citeauthoryear{{Popov} and {Prokhorov}}{2007}]{pp2007}
\begin{barticle}
\bauthor{\bsnm{{Popov}}, \binits{S.B.}},
\bauthor{\bsnm{{Prokhorov}}, \binits{M.E.}}:
\bjtitle{Physics Uspekhi}
\bvolume{50},
\bfpage{1123}
(\byear{2007}).
\arxivurl{arXiv:astro-ph/0411792}.
doi:\doiurl{10.1070/PU2007v050n11ABEH006179}
\end{barticle}
\endbibitem

\bibitem[\protect\citeauthoryear{{Popov} et~al.}{2010}]{popov2010}
\begin{barticle}
\bauthor{\bsnm{{Popov}}, \binits{S.B.}},
\bauthor{\bsnm{{Pons}}, \binits{J.A.}},
\bauthor{\bsnm{{Miralles}}, \binits{J.A.}},
\bauthor{\bsnm{{Boldin}}, \binits{P.A.}},
\bauthor{\bsnm{{Posselt}}, \binits{B.}}:
\bjtitle{\mnras}
\bvolume{401},
\bfpage{2675}
(\byear{2010}).
\arxivurl{0910.2190}.
doi:\doiurl{10.1111/j.1365-2966.2009.15850.x}
\end{barticle}
\endbibitem

\bibitem[\protect\citeauthoryear{{Porquet} et~al.}{2003}]{porquet2003}
\begin{barticle}
\bauthor{\bsnm{{Porquet}}, \binits{D.}},
\bauthor{\bsnm{{Decourchelle}}, \binits{A.}},
\bauthor{\bsnm{{Warwick}}, \binits{R.S.}}:
\bjtitle{\aap}
\bvolume{401},
\bfpage{197}
(\byear{2003}).
\arxivurl{arXiv:astro-ph/0211426}.
doi:\doiurl{10.1051/0004-6361:20021670}
\end{barticle}
\endbibitem

\bibitem[\protect\citeauthoryear{{Reynolds}}{1985}]{reynolds1985}
\begin{barticle}
\bauthor{\bsnm{{Reynolds}}, \binits{S.P.}}:
\bjtitle{\apj}
\bvolume{291},
\bfpage{152}
(\byear{1985}).
doi:\doiurl{10.1086/163050}
\end{barticle}
\endbibitem

\bibitem[\protect\citeauthoryear{{Roberts} and {Brogan}}{2008}]{g7.5-1.7}
\begin{barticle}
\bauthor{\bsnm{{Roberts}}, \binits{M.S.E.}},
\bauthor{\bsnm{{Brogan}}, \binits{C.L.}}:
\bjtitle{\apj}
\bvolume{681},
\bfpage{320}
(\byear{2008}).
\arxivurl{0802.3750}.
doi:\doiurl{10.1086/588419}
\end{barticle}
\endbibitem

\bibitem[\protect\citeauthoryear{{Rudie} and {Fesen}}{2007}]{rudie2007}
\begin{bchapter}
\bauthor{\bsnm{{Rudie}}, \binits{G.C.}},
\bauthor{\bsnm{{Fesen}}, \binits{R.A.}}:
In: \bbtitle{Revista Mexicana de Astronomia y Astrofisica Conference Series}.
\bsertitle{Revista Mexicana de Astronomia y Astrofisica, vol. 27},
vol. \bseriesno{30},
p. \bfpage{90}
(\byear{2007}).
\arxivurl{0704.2780}
\end{bchapter}
\endbibitem

\bibitem[\protect\citeauthoryear{{Safi-Harb} et~al.}{2001}]{g21.5-0.9}
\begin{barticle}
\bauthor{\bsnm{{Safi-Harb}}, \binits{S.}},
\bauthor{\bsnm{{Harrus}}, \binits{I.M.}},
\bauthor{\bsnm{{Petre}}, \binits{R.}},
\bauthor{\bsnm{{Pavlov}}, \binits{G.G.}},
\bauthor{\bsnm{{Koptsevich}}, \binits{A.B.}},
\bauthor{\bsnm{{Sanwal}}, \binits{D.}}:
\bjtitle{\apj}
\bvolume{561},
\bfpage{308}
(\byear{2001}).
\arxivurl{arXiv:astro-ph/0107175}.
doi:\doiurl{10.1086/322978}
\end{barticle}
\endbibitem

\bibitem[\protect\citeauthoryear{{Scheck} et~al.}{2006}]{scheck2006}
\begin{barticle}
\bauthor{\bsnm{{Scheck}}, \binits{L.}},
\bauthor{\bsnm{{Kifonidis}}, \binits{K.}},
\bauthor{\bsnm{{Janka}}, \binits{H.-T.}},
\bauthor{\bsnm{{M{\"u}ller}}, \binits{E.}}:
\bjtitle{\aap}
\bvolume{457},
\bfpage{963}
(\byear{2006}).
\arxivurl{arXiv:astro-ph/0601302}.
doi:\doiurl{10.1051/0004-6361:20064855}
\end{barticle}
\endbibitem

\bibitem[\protect\citeauthoryear{{Slane} et~al.}{2004}]{Slane2004}
\begin{barticle}
\bauthor{\bsnm{{Slane}}, \binits{P.}},
\bauthor{\bsnm{{Zimmerman}}, \binits{E.R.}},
\bauthor{\bsnm{{Hughes}}, \binits{J.P.}},
\bauthor{\bsnm{{Seward}}, \binits{F.D.}},
\bauthor{\bsnm{{Gaensler}}, \binits{B.M.}},
\bauthor{\bsnm{{Clarke}}, \binits{M.J.}}:
\bjtitle{\apj}
\bvolume{601},
\bfpage{1045}
(\byear{2004}).
\arxivurl{arXiv:astro-ph/0310250}.
doi:\doiurl{10.1086/380498}
\end{barticle}
\endbibitem

\bibitem[\protect\citeauthoryear{{Slane} et~al.}{2008}]{3c58}
\begin{barticle}
\bauthor{\bsnm{{Slane}}, \binits{P.}},
\bauthor{\bsnm{{Helfand}}, \binits{D.J.}},
\bauthor{\bsnm{{Reynolds}}, \binits{S.P.}},
\bauthor{\bsnm{{Gaensler}}, \binits{B.M.}},
\bauthor{\bsnm{{Lemiere}}, \binits{A.}},
\bauthor{\bsnm{{Wang}}, \binits{Z.}}:
\bjtitle{\apjl}
\bvolume{676},
\bfpage{33}
(\byear{2008}).
\arxivurl{0802.0206}.
doi:\doiurl{10.1086/587031}
\end{barticle}
\endbibitem

\bibitem[\protect\citeauthoryear{{Stephenson} and {Green}}{2002}]{snr2002}
\begin{botherref}
\oauthor{\bsnm{{Stephenson}}, \binits{F.R.}},
\oauthor{\bsnm{{Green}}, \binits{D.A.}}:
Historical supernovae and their remnants, by F.~Richard Stephenson and David
  A.~Green.~International series in astronomy and astrophysics, vol.~5.~Oxford:
  Clarendon Press, 2002, ISBN 0198507666
\textbf{5}
(2002)
\end{botherref}
\endbibitem

\bibitem[\protect\citeauthoryear{{Sun} et~al.}{2004}]{kes79}
\begin{barticle}
\bauthor{\bsnm{{Sun}}, \binits{M.}},
\bauthor{\bsnm{{Seward}}, \binits{F.D.}},
\bauthor{\bsnm{{Smith}}, \binits{R.K.}},
\bauthor{\bsnm{{Slane}}, \binits{P.O.}}:
\bjtitle{\apj}
\bvolume{605},
\bfpage{742}
(\byear{2004}).
\arxivurl{arXiv:astro-ph/0401165}.
doi:\doiurl{10.1086/382666}
\end{barticle}
\endbibitem

\bibitem[\protect\citeauthoryear{{Thorsett} et~al.}{2003}]{monogem}
\begin{barticle}
\bauthor{\bsnm{{Thorsett}}, \binits{S.E.}},
\bauthor{\bsnm{{Benjamin}}, \binits{R.A.}},
\bauthor{\bsnm{{Brisken}}, \binits{W.F.}},
\bauthor{\bsnm{{Golden}}, \binits{A.}},
\bauthor{\bsnm{{Goss}}, \binits{W.M.}}:
\bjtitle{\apjl}
\bvolume{592},
\bfpage{71}
(\byear{2003}).
\arxivurl{arXiv:astro-ph/0306462}.
doi:\doiurl{10.1086/377682}
\end{barticle}
\endbibitem

\bibitem[\protect\citeauthoryear{{Tian} and {Leahy}}{2006}]{g65.1+0.6}
\begin{barticle}
\bauthor{\bsnm{{Tian}}, \binits{W.W.}},
\bauthor{\bsnm{{Leahy}}, \binits{D.A.}}:
\bjtitle{\aap}
\bvolume{455},
\bfpage{1053}
(\byear{2006}).
\arxivurl{arXiv:astro-ph/0603102}.
doi:\doiurl{10.1051/0004-6361:20065140}
\end{barticle}
\endbibitem

\bibitem[\protect\citeauthoryear{{Torii} et~al.}{1999}]{g11.2-0.3}
\begin{barticle}
\bauthor{\bsnm{{Torii}}, \binits{K.}},
\bauthor{\bsnm{{Tsunemi}}, \binits{H.}},
\bauthor{\bsnm{{Dotani}}, \binits{T.}},
\bauthor{\bsnm{{Mitsuda}}, \binits{K.}},
\bauthor{\bsnm{{Kawai}}, \binits{N.}},
\bauthor{\bsnm{{Kinugasa}}, \binits{K.}},
\bauthor{\bsnm{{Saito}}, \binits{Y.}},
\bauthor{\bsnm{{Shibata}}, \binits{S.}}:
\bjtitle{\apjl}
\bvolume{523},
\bfpage{69}
(\byear{1999}).
doi:\doiurl{10.1086/312251}
\end{barticle}
\endbibitem

\bibitem[\protect\citeauthoryear{{Uchiyama} et~al.}{2002}]{g78.2+2.1}
\begin{barticle}
\bauthor{\bsnm{{Uchiyama}}, \binits{Y.}},
\bauthor{\bsnm{{Takahashi}}, \binits{T.}},
\bauthor{\bsnm{{Aharonian}}, \binits{F.A.}},
\bauthor{\bsnm{{Mattox}}, \binits{J.R.}}:
\bjtitle{\apj}
\bvolume{571},
\bfpage{866}
(\byear{2002}).
\arxivurl{arXiv:astro-ph/0202414}.
doi:\doiurl{10.1086/340121}
\end{barticle}
\endbibitem

\bibitem[\protect\citeauthoryear{{Vasisht} et~al.}{1996}]{vasisht1996}
\begin{barticle}
\bauthor{\bsnm{{Vasisht}}, \binits{G.}},
\bauthor{\bsnm{{Aoki}}, \binits{T.}},
\bauthor{\bsnm{{Dotani}}, \binits{T.}},
\bauthor{\bsnm{{Kulkarni}}, \binits{S.R.}},
\bauthor{\bsnm{{Nagase}}, \binits{F.}}:
\bjtitle{\apjl}
\bvolume{456},
\bfpage{59}
(\byear{1996}).
doi:\doiurl{10.1086/309854}
\end{barticle}
\endbibitem

\bibitem[\protect\citeauthoryear{{Vasisht} et~al.}{1997}]{g296.5+10.0}
\begin{barticle}
\bauthor{\bsnm{{Vasisht}}, \binits{G.}},
\bauthor{\bsnm{{Kulkarni}}, \binits{S.R.}},
\bauthor{\bsnm{{Anderson}}, \binits{S.B.}},
\bauthor{\bsnm{{Hamilton}}, \binits{T.T.}},
\bauthor{\bsnm{{Kawai}}, \binits{N.}}:
\bjtitle{\apjl}
\bvolume{476},
\bfpage{43}
(\byear{1997}).
doi:\doiurl{10.1086/310493}
\end{barticle}
\endbibitem

\bibitem[\protect\citeauthoryear{{Wang} and {Gotthelf}}{1998}]{wg1998}
\begin{barticle}
\bauthor{\bsnm{{Wang}}, \binits{Q.D.}},
\bauthor{\bsnm{{Gotthelf}}, \binits{E.V.}}:
\bjtitle{\apj}
\bvolume{494},
\bfpage{623}
(\byear{1998}).
\arxivurl{arXiv:astro-ph/9708087}.
doi:\doiurl{10.1086/305214}
\end{barticle}
\endbibitem

\bibitem[\protect\citeauthoryear{{Williams} et~al.}{2008}]{williams2008}
\begin{barticle}
\bauthor{\bsnm{{Williams}}, \binits{B.J.}},
\bauthor{\bsnm{{Borkowski}}, \binits{K.J.}},
\bauthor{\bsnm{{Reynolds}}, \binits{S.P.}},
\bauthor{\bsnm{{Raymond}}, \binits{J.C.}},
\bauthor{\bsnm{{Long}}, \binits{K.S.}},
\bauthor{\bsnm{{Morse}}, \binits{J.}},
\bauthor{\bsnm{{Blair}}, \binits{W.P.}},
\bauthor{\bsnm{{Ghavamian}}, \binits{P.}},
\bauthor{\bsnm{{Sankrit}}, \binits{R.}},
\bauthor{\bsnm{{Hendrick}}, \binits{S.P.}},
\bauthor{\bsnm{{Smith}}, \binits{R.C.}},
\bauthor{\bsnm{{Points}}, \binits{S.}},
\bauthor{\bsnm{{Winkler}}, \binits{P.F.}}:
\bjtitle{\apj}
\bvolume{687},
\bfpage{1054}
(\byear{2008}).
\arxivurl{0807.4155}.
doi:\doiurl{10.1086/592139}
\end{barticle}
\endbibitem

\bibitem[\protect\citeauthoryear{{Winkler} et~al.}{1988}]{wtki1988}
\begin{bchapter}
\bauthor{\bsnm{{Winkler}}, \binits{P.F.}},
\bauthor{\bsnm{{Tuttle}}, \binits{J.H.}},
\bauthor{\bsnm{{Kirshner}}, \binits{R.P.}},
\bauthor{\bsnm{{Irwin}}, \binits{M.J.}}:
In: \beditor{\bsnm{{R.~S.~Roger \& T.~L.~Landecker}}} (ed.)
\bbtitle{IAU Colloq. 101: Supernova Remnants and the Interstellar Medium},
p. \bfpage{65}
(\byear{1988})
\end{bchapter}
\endbibitem

\bibitem[\protect\citeauthoryear{{Yar-Uyaniker} et~al.}{2004}]{g114.3+0.3}
\begin{barticle}
\bauthor{\bsnm{{Yar-Uyaniker}}, \binits{A.}},
\bauthor{\bsnm{{Uyaniker}}, \binits{B.}},
\bauthor{\bsnm{{Kothes}}, \binits{R.}}:
\bjtitle{\apj}
\bvolume{616},
\bfpage{247}
(\byear{2004}).
\arxivurl{arXiv:astro-ph/0408386}.
doi:\doiurl{10.1086/424794}
\end{barticle}
\endbibitem

\bibitem[\protect\citeauthoryear{{Yatsu} et~al.}{2005}]{g320.4-1.2}
\begin{barticle}
\bauthor{\bsnm{{Yatsu}}, \binits{Y.}},
\bauthor{\bsnm{{Kataoka}}, \binits{J.}},
\bauthor{\bsnm{{Kawai}}, \binits{N.}},
\bauthor{\bsnm{{Tamura}}, \binits{K.}},
\bauthor{\bsnm{{Brinkmann}}, \binits{W.}}:
\bjtitle{Advances in Space Research}
\bvolume{35},
\bfpage{1066}
(\byear{2005}).
doi:\doiurl{10.1016/j.asr.2005.05.015}
\end{barticle}
\endbibitem

\bibitem[\protect\citeauthoryear{{Zhang} and {Xie}}{2011}]{china}
\begin{botherref}
\oauthor{\bsnm{{Zhang}}, \binits{S.-N.}},
\oauthor{\bsnm{{Xie}}, \binits{Y.}}:
ArXiv e-prints
(2011).
\arxivurl{1110.3154}
\end{botherref}
\endbibitem

\bibitem[\protect\citeauthoryear{{Zhang} and {Xie}}{2012}]{xie2012}
\begin{botherref}
\oauthor{\bsnm{{Zhang}}, \binits{S.-N.}},
\oauthor{\bsnm{{Xie}}, \binits{Y.}}:
ArXiv e-prints
(2012).
\arxivurl{1202.1123}
\end{botherref}
\endbibitem

\end{thebibliography}

\end{document}